\documentclass{ifacconf}

\usepackage{graphicx}
\usepackage{epstopdf}      % include this line if your document contains figures
\usepackage{natbib}        % required for bibliography
\usepackage{tikz}
\usepackage{subfig}
\usetikzlibrary{decorations.shapes,shapes.geometric,shadows,arrows,automata,positioning,calendar,mindmap,backgrounds,scopes,chains,er,3d,matrix,fit}
\usetikzlibrary{shapes.symbols}
\usepackage{amssymb}  % assumes amsmath package installed
\usepackage{mathdots}% http://ctan.org/pkg/mathdots
\usepackage{algorithm}
\usepackage{algpseudocode}
\usetikzlibrary{fit}
\usepackage{amsmath} % assumes amsmath package installed
\usepackage{balance}
\usepackage{amsmath}
\usepackage{multicol}
\usepackage{graphicx}
\usepackage{mathrsfs}
\newcommand{\R}{\mathbb{R}}
\newcommand{\B}{\mathbb{B}}

\newcommand{\cP}{\mathcal{P}}%

\newcommand{\cS}{\mathcal{S}}
\newcommand{\cU}{\mathcal{U}}
\newcommand{\cX}{\mathcal{X}}

\newcommand{\cC}{\mathcal{C}}
\newcommand{\cE}{\mathcal{E}}

\newcommand{\cB}{\mathcal{B}}

\newcommand{\cV}{\mathcal{V}}
\newcommand{\cF}{\mathcal{F}}
\newcommand{\cR}{\mathcal{R}}
\newcommand{\cH}{\mathcal{H}}
\newcommand{\cZ}{\mathcal{Z}}

\newcommand{\cW}{\mathcal{W}}
\newcommand{\cY}{\mathcal{Y}}
\newcommand{\cG}{\mathcal{G}}
\newcommand{\cA}{\mathcal{A}}

\newcommand{\conv}[1]{\operatorname{conv}({#1})}

\newcommand{\Out}[2]{\operatorname{Out}(#1,\;#2)}
\newcommand{\interior}{\operatorname{int}}

\providecommand{\com[2]}{\begin{tt}[#1: #2]\end{tt}}

\newcommand{\vol}{\operatorname{vol}}
\usepackage{color}

\newcommand{\Red}[1]{{\color{red}{#1}}}
\newcommand{\Blue}[1]{{\color{blue}{#1}}}

\begin{document}
\begin{frontmatter}

\title{Quantifying impact on safety from cyber-attacks on cyber-physical systems\thanksref{footnoteinfo}} 

\thanks[footnoteinfo]{ E.V. and N.A. gratefully acknowledge support from EPSRC EP/T021942/1, and N.A. additionally from EU 2020-1-UK01-KA203-079283 and the UKRI Belfast Maritime Consortium 107138.}

\author[First]{Eleftherios Vlahakis} 
\author[Second]{Gregory Provan} 
\author[Third]{Gordon Werner}
\author[Third]{Shanchieh Yang}
\author[First]{Nikolaos Athanasopoulos}

\address[First]{Queen's University Belfast, UK (e-mail: \{e.vlahakis, n.athanasopoulos\}@qub.ac.uk).}
\address[Second]{University College Cork, Ireland (e-mail: g.provan@cs.ucc.ie)}
\address[Third]{Rochester Institute of Technology, Rochester, USA, (e-mail: gxw9834@rit.edu, jay.yang@rit.edu)}

\begin{abstract}                % Abstract of not more than 250 words.
We propose a novel framework for modelling attack scenarios in cyber-physical control systems: we represent a cyber-physical system as a constrained switching system, where a single model embeds the dynamics of the physical process, the attack patterns, and the attack detection schemes. We show that this is compatible with established results in the analysis of hybrid automata, and, specifically, constrained switching systems. Moreover, we use the developed models to compute the impact of cyber attacks on the safety properties of the system. In particular, we characterise system safety as an asymptotic property, by calculating the maximal safe set. The resulting new impact metrics intuitively quantify the degradation of safety under attack. We showcase our results via illustrative examples.
\end{abstract}


\end{frontmatter}
%===============================================================================

\section{Introduction}

Cyber-physical systems (CPSs) can represent a broad spectrum of safety-critical applications, ranging from power generation and distribution networks to autonomous mobility and industrial processes. Due to their extent and intrinsic link to society,  secure operation of such schemes is vital. Vulnerability to cyber attacks typically depends on the degree of integrating unsafe communication channels between computation, sensing, and actuation modules that control the underlying physical process.

CPS security research, e.g., \citep{Sandberg2022, Milosevic2020}, studies control problems under adversarial actions, which aim to steer a control system into an unsafe region. Our modelling approach is motivated to an extent by the literature of networked control systems (NCSs), \citep{Hespanha2007}, where typical communication limitations and malfunctions, such as sampling, network delay, and packet dropouts, can be embedded into a hybrid control system model with switching dynamics, see, e.g., \citep{HeemelsTAC2011, DePersisTAC2015, Zhang2008}. Due to their inherent complexity, CPS are often modelled via hybrid systems. For example, hybrid linear automata, finite state machines, and Petri nets are important tools for modelling malicious and unpredictable behaviours, and threat propagations in CPSs \citep{Davoudi2017, Lafortune2020, Liu2017}. 

Focusing on attack patterns and logic rules that can be expressed via regular languages on directed labelled graphs \citep{CassandrasBook2010}, we propose a constrained switching systems framework for analysing safety properties of CPSs. Although invariance and safety of constrained switching systems, \citep{AthanasopoulosCDC14, Dai2012, Philippe2016} have received attention \citep{AthanasopoulosLCSS2017, Athanasopoulos2018, DeSantisTAC2004}, they have not been studied yet in the context of security of CPS. By modelling the overall attack scheme as a constrained switching system, our objective is to characterise the set of all initial states that cannot be driven to an unsafe state under any allowable attack. We call this the \emph{safe set} of the attacked CPS. In the absence of switching dynamics, this is an infinite-reachability, dynamic programming problem \citep{RakovicTAC2006, BlanchiniBook2015}: The maximal safe set can be retrieved by computing, in a recursive fashion, the fixed point of the sequence of sets $\{S_i\}_{i \in \{1,2,\ldots\}}$ with
$S_{i+1} = \textnormal{Pre}(S_i) \cap S_0,$ where $S_0 = X_0$ denotes the state-constraints set, and $\textnormal{Pre}(S_i)$ is the \emph{preimage map} \citep{BertsekasTAC1972}, that is the set of states $x$ for which, for all permissible attack patterns, the successor state $x^+\in S_i$.

There is significant ongoing research on modeling CPS security,  with many challenges still remaining. One challenge concerns analysis of temporal aspects of attacks, and the use of such models to assess system safety. Some prior work has adopted automata as the language for attack models. These automata do not define timing behaviours, nor do they discuss taking cross-products of separate attacks to specify an automaton for multiple simultaneous attacks. \citep{chen2003data} proposes a data-driven Finite State Machine model for analyzing security vulnerabilities. \citep{zhang2012research} defines an alternative view of attack modeling, as based on Finite State Machines. \citep{james2021situational} defines a methodology for Finite State Automata-based attack modeling, as applied to situational awareness for smart home IoT security. \citep{reda2022comprehensive} surveys the state-or-the-art in  false data injection attacks in smart grids, focusing on attack models, targets, and impacts. Few papers have introduced general languages for describing attacks models. \citep{Liu2017} has proposed a language based on probabilistic colored Petri nets, which together with mixed-strategy game theory is used for modeling cyber-physical attacks. This is the most comprehensive formal language yet specified for this purpose.
%---------

Open high-impact security issues in modeling include specifying attacks being persistent or intermittent. The ability to formally model attacks and compose such models to describe multiple simultaneous attacks also has received little attention. Most importantly, most works derive receding horizon impact metrics of attacks on the state space of the closed-loop system. Game-theoretic approaches (e.g., \citep{Zhu2015, ZhuCDC2014}) provide meaningful  answers to the above challenges, however are concerned to a finite set of outcomes (e.g., best-case or worst-case scenarios), and are concerned with exploring the space of possible safety violations. Our approach is aligned with the reachability-analysis-based works \citep{Nesic2020, SinopoliTAC2016}, providing asymptotic results on the safety of the closed-loop system. We treat stealthy attack perturbations as state- and/or input-dependent exogenous signals  \citep{RakovicTAC2006, CannonCDC15}. To our knowledge, there is very limited work dealing with malicious state-dependent attacks directly in the context of CPSs.  In this work, we focus on stealthy false data injection (FDI) attacks, and follow standard modelling techniques primarily motivated by the approach in \citep{Teixeira2015}. Our contributions can be summarised as follows:

\noindent $\bullet$ We model the overall CPS under attack as a constrained switching system with the switching signal forming a regular language, generated by a nondeterministic directed graph. Each node of the graph is associated with a set of states that evolve with time according to the modes assigned to the corresponding outgoing edges. Each labelled edge describes either an attack-free operation or a specific malicious action carried out over a subset of unsafe channels. This approach to attack modelling gives us considerable flexibility in modelling a large class of non-deterministic attack patterns.

\noindent $\bullet$ We propose a new approach of quantifying the asymptotic impact of attacks on the system via the construction of maximal safe sets.
To compute the maximal safe set of the system subject to all admissible attack sequences, we leverage reachability analysis techniques related to the notion of multi-set invariance, \citep{AthanasopoulosLCSS2017, Athanasopoulos2018}. Based on the constructed sets, we assess vulnerability by two complementary security metrics, related to the Lebesgue measure and the Minkowski distance,  providing   scalar indices of system attack sensitivity. 

The remainder of the paper is organised as follows. In Section \ref{sec:system}, we present the family of systems we study and the type of attacks we are interested in. The main results, namely, the constrained switching system formulation, the safe set computations, and the introduction of scalar safety metrics, are in Section \ref{section:switching_sys_model}. A numerical example and concluding remarks are in Sections \ref{sec:example} and \ref{sec:conclusion}, respectively.

\section{System description}\label{sec:system}

We present the formal dynamic system model with the associated control and estimator schemes. We introduce the interplay of the nominal system with the malicious signals and define the anomaly detection unit. We consider the outputs of the sensors and the controller as the vulnerable points. See Fig. \ref{fig:1} for an illustration.

\subsection{Notation}
The set of nonnegative real and natural numbers is $\mathbb{R}_+$ and $\mathbb{N}$ respectively. $\mathbb{R}^{n}$ denotes the real $n$-dimensional vector space, and $\mathbb{R}^{n\times m}$ denotes the set of $n\times m$ real matrices. The transpose of a vector $\xi$ is $\xi^\top$. The $m\times m$ identity matrix is $I_{m}$ and the vector with elements equal to one is $1\in\mathbb{R}^n$. The $j$th row of matrix $A$ and $j$-th element of  vector $a$ are denoted by $(A)_j$ and $(a)_j$, respectively. The set of row indices of $A$ is $J_A$. We write $\cG(\cV,\cE)$, or $\cG$, a labelled directed graph with a set of nodes $\cV$ and a set of edges $\cE$. We denote the $p$-norm of a vector $x$ by $\|x\|_p$, and the vector with all of its elements equal to one by $1$. $\B(\alpha)$, and $\B_{\infty}(\alpha)$ denote the balls of radius $\alpha$ of an arbitrary norm, and the infinity norm, respectively. The Minkowski sum of two sets $\cS_1$ and $\cS_2$ is denoted by  $\cS_1\oplus\cS_2$. The interior and the convex hull of a set $\cS$ are denoted as $\interior{(\cS)}$ and $\conv{\cS}$, respectively. A C-set $\cS\subset\R^n$ is a convex compact polytopic set which contains the origin in its interior, \citep{BlanchiniBook2015}. By convention, for any $C$-set  $\cV$, we write its half-space representation by $\cV = \{v: G_v v \leq g_v\}$ with the inequalities applying elementwise. The cardinality of a set $\cV$ is denoted by $|\cV|$.

\subsection{Dynamics} 

We study discrete-time linear time-invariant (LTI) systems
\begin{equation}
 P:   \Bigg\{\begin{aligned}\label{eq:dynamics}
    x(t+1) &= A_p x(t) + B_pu(t) + v(t),\\
    y(t) &= C_p x(t) + w(t),
    \end{aligned}
\end{equation}
where $t\in \mathbb{N}$, $x(t) \in \cX \subset \mathbb{R}^{n_x}$, $u(t)\in \cU\subset \mathbb{R}^{n_u}$ and $y(t)\in \cY\subset\mathbb{R}^{n_y}$ are the state, input and output vectors, respectively, vectors $v(t)\in\cV\subset\R^{n_x}$  and $w(t)\in\cW\subset\R^{n_y}$ denote process and measurement uncertainties. We assume that $\cX$, $\cU$, $\cY$, $\cV$, and $\cW$ are $C$-sets. For a meaningful control and estimation scheme, we assume the following.
\begin{assum}\label{ass:stabilisability}
The pairs $(A_p,\;B_p)$, $(A_p^\top,\;C_p^\top)$ are stabilisable.
\end{assum}

\subsection{Dynamic output feedback}

\begin{figure}[ht] 
\begin{center}
\begin{tikzpicture}[->,>=stealth',shorten >=1pt,auto,node distance=1.6cm, font=\small, semithick]
\tikzstyle{every state}=[fill=gray!20,draw=white, text=black,scale=.8]
\draw[solid] (0,0) -- (1,0);
\node[draw, fill = gray!20, fit={(0,-.35) (1.5,.35)}, xshift = 0cm, inner sep=0pt, label=center:Actuators] (A) {};
\node[draw, fill = gray!20, fit={(0,-.5) (1.5,.5)}, xshift = 3cm, inner sep=0pt, label=center:Process] (B) {};
\node[draw, fill = gray!20, fit={(0,-.35) (1.5,.35)}, xshift = 6cm, inner sep=0pt, label=center:Sensors] (C) {};
\node (D) [above of = B] {};
\node (E) [below of = D, yshift = .9cm, xshift = -.5cm] {$v(t)$};
\coordinate (EE) at at ([yshift={.9cm}]B);
\draw (A)--(B);
\draw (B)--(C);
\draw (E)--(EE)--(B);
\node (F) [above of = C] {};
\node (G) [below of = F, yshift = .75cm, xshift = -.5cm] {$w(t)$};
\coordinate (GG) at at ([yshift={.75cm}]C);
\draw (G)--(GG)--(C);
\node[state, draw = gray!90, dashed] (H) [below of = A] {$\scalebox{3}{+}$};
\path[<-,dashed, gray!90, text = black] (A) edge node[right] {$\tilde{u}(t)$} (H);
\node (I) [right of = H] {};
\draw[red,->] (I)--node[above] {$a_u(t)$}(H);
\coordinate (K) at ([yshift={-1.cm}]H);
\coordinate (L) at ([xshift = {1.1cm}]K);
\draw[<-,dashed, gray!90, text = black] (H)--(K)--node[above] {$u(t)$}(L);
\node[draw, fill = gray!20, fit={(0,-.35) (1.5,.35)}, right of = K, xshift = .25cm, inner sep=0pt, label=center:Controller] (N) {};
\node[draw, fill = gray!20, fit={(0,-.35) (1.5,.35)}, right of = N, xshift = .8cm, inner sep=0pt, label=center:Estimator] (O) {};
\path[<-] (N) edge  node[above] {$\hat{x}(t)$} (O);
\node[state, draw = gray!90, dashed] (P) [below of = C] {$\scalebox{3}{+}$};
\path[->,dashed, gray!90, text = black] (C) edge node[right] {$y(t)$} (P);
\node (II) [left of = P] {};
\draw[red,->] (II)--node[above] {$a_y(t)$}(P);
\coordinate (KK) at ([yshift={-1.cm}]P);
\coordinate (KKK) at ([yshift={-.9cm}]KK);
% \path[-,dashed, gray!90] (K) edge (H);
\coordinate (LL) at ([xshift = {-1cm}]KK);
\draw[->,dashed, gray!90, text = black] (P)--node[right] {$\tilde{y}(t)$}(KK)--(LL);
\coordinate (NN) at ([yshift={1.cm}]N);
\coordinate (OO) at ([yshift={1.cm}]O);
\draw[->] (N)--(NN)--node[above] {$u(t)$}(OO)--(O);
\coordinate (DD) at ([xshift={1.2cm}]N);
\node[draw, fill = gray!20, fit={(0,-.35) (1.5,.35)}, below of = DD, xshift = .cm, yshift = .7cm, inner sep=0pt, label=center:Detector] (DDD) {};
\draw[->] (DD)--(DDD);
\draw[->, dashed, gray!90] (P)--(KK)--(KKK)--(DDD);
\end{tikzpicture}
\end{center}
\caption{Networked control loop with sensor and actuation poisoning attacks. Dashed gray lines denote unsafe communication channels. Solid black lines denote secure or physical connections.}
\label{fig:1}
\end{figure}
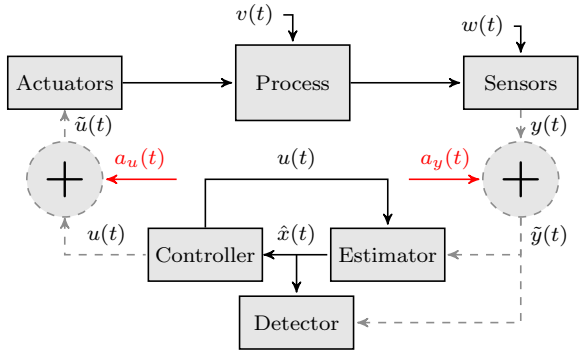

We are interested in \emph{false data injection} (FDI) attacks, namely, sensor poisoning and input poisoning attacks associated with $a_y(t)$ and $a_u(t)$ in Fig. \ref{fig:1}, respectively. 

The sensor output $y(t)$ is prone to corruption, and we model the attacked output as 
\begin{equation}\label{eq:attacked_output}
    \tilde{y}(t) = y(t) + \Gamma_i^y a_y(t),
\end{equation}
where $a_y(t)\in \mathbb{R}^{n_{\tilde{y}}}$ denotes \emph{additive sensor poisoning attacks}, and $\Gamma_i^y \in \mathbb{R}^{n_y \times n_{\tilde{y}}}$, with $n_{\tilde{y}} \leq n_y$ denoting the number of vulnerable sensors. The subscript $i$ denotes the $i$th attack strategy.\footnote{See Section \ref{sec:attack}.} The $j$th row of $\Gamma_i^y$ is $0\in\mathbb{R}^{1\times n_{\tilde{y}}}$ if the $j$th sensor is not corrupted under the $i$th attack action. Otherwise, it is the $\tilde{j}$th vector $\epsilon_{\tilde{j}}$ of the canonical basis of $\mathbb{R}^{n_{\tilde{y}}\times n_{\tilde{y}}}$, with $\tilde{j}$ denoting the index of a vulnerable sensor under attack by the $i$th attack action.

The vector $\tilde{y}(t)$ is received by the controller. We consider a stabilising dynamic output feedback control law   
\begin{equation}\label{eq:controller}
    u(t) = -K\hat{x}(t),
\end{equation}
with $K\in \mathbb{R}^{n_u\times n_x}$. The signal $\hat{x}(t)$ is an estimate of $x(t)$ obtained by the estimator
\begin{equation}\label{eq:estimator}
    \hat{x}(t+1) = A_p\hat{x}(t) + B_pu(t) + L(\tilde{y}(t) - C_p\hat{x}(t)),
\end{equation}
with observer gain $L\in \mathbb{R}^{n_x \times n_y}$. We call $r(t)=\tilde{y}(t) - C_p\hat{x}(t)\in\mathbb{R}^{n_y}$ the \emph{residual} and $e(t) = x(t) - \hat{x}(t)$ the \emph{estimation error}. From \eqref{eq:dynamics}-\eqref{eq:estimator}, we may write the dynamics of $e(t)$ and $r(t)$ as
\begin{equation}\label{eq:error_residual}
    \Bigg\{\begin{aligned}
    e(t+1) &= (A_p-LC_p)e(t) -L \Gamma_i^y a_y(t) -Lw(t)  \\
    r(t) &= C_pe(t) + \Gamma_i^y a_y(t) + w(t). 
\end{aligned}  
\end{equation}

\begin{rem}
Under Assumption \ref{ass:stabilisability}, the controller and observer gains, $K$ and $L$, respectively, are designed such that $A_p-B_pK$ and $A_p-LC_p$ are Schur matrices. A desirable robust performance for the attack-free system \eqref{eq:dynamics} can be achieved by constructing $K$ and $L$ via  LMI-based algorithms \citep{Gahinet1994}. 
\end{rem}

The received control signal is corrupted as 
\begin{equation}\label{eq:attacked_control}
    \tilde{u}(t) = u(t) + \Gamma_i^u a_u(t),
\end{equation}
where $a_u(t) \in \mathbb{R}^{n_{\tilde{u}}}$ denotes \emph{additive input poisoning attacks}, $\Gamma_i^u \in \mathbb{R}^{n_u\times n_{\tilde{u}}}$, with $n_{\tilde{u}} \leq n_u$ denoting the number of unsafe channels over which actuation signals are transmitted. The structure of $\Gamma_i^u$ is associated with the $i$th attack action and is in line with the one of $\Gamma_i^u$ defined in \eqref{eq:attacked_output}.

\subsection{Detector}

We consider that a nominal (fault-free) operation is attained if $r(t)\in \cR$ with
\begin{equation}\label{eq:stealthy_residual_set}
    \cR = \{ r\in\mathbb{R}^{n_y}:\; G_r r \leq h_r \},
\end{equation}
where $G_r$ and $w_r$ are of appropriate dimensions, with $G_r$ being a full row-rank matrix. An alarm is raised at $t\geq 0$ if $r(t)\notin \cR$.
\begin{rem}
The polyhedral set $\cR$ may be designed such that the number of false alarms is minimised subject to the process and measurement perturbations.
\end{rem}
\begin{rem}
Our anomaly monitoring scheme is a \emph{stateless} detector. \emph{Stateful} detectors with linear, or convex, dynamics can be accepted in our framework, \citep{MilosevicECC18}. 
\end{rem}

\subsection{Closed-loop dynamics}

The closed-loop dynamics can be written in terms of both the states $x(t)$ and the estimation error $e(t)$. By defining the augmented vectors $z(t) =[x(t)^{\top}\;e(t)^{\top}]^{\top}$, $a(t) = [a_u(t)^{\top},\;a_y(t)^{\top}]^{\top}$, and $\eta(t) = [v(t)^{\top},\;w(t)^{\top}]^{\top}$, we write the closed-loop dynamics under the $i$th attack action as
\begin{equation}\label{eq:closed_loop2}
    P_i:    \Bigg\{ \begin{aligned}
        z(t+1) = Az(t) + B_ia(t) + E\eta(t),\\
        r(t) = Cz(t) + D_ia(t) + F\eta(t),
    \end{aligned}
\end{equation}
where
\begin{align*}
    & A = \begin{bmatrix} A_p -B_pK & B_pK
    \\0_{n_x\times n_x} & A_p -LC_p \end{bmatrix},\;B_i = \begin{bmatrix} B_p\Gamma_i^u & 0_{n_x\times n_{\tilde{y}}}  \\ 0_{n_x\times n_{\tilde{u}}} & -L\Gamma_i^y \end{bmatrix}  \\ 
    & E = \begin{bmatrix} I_{n_x} & 0_{n_x\times n_y}\\ I_{n_x} & -L \end{bmatrix},\; C = \begin{bmatrix} 0_{n_y \times n_x} & C_p \end{bmatrix}, \\ & D_i = \begin{bmatrix} 0_{n_y\times n_{\tilde{u}}} & \Gamma_i^y \end{bmatrix},\; F = \begin{bmatrix} 0_{n_y\times n_x} & I_{n_y} \end{bmatrix}. 
\end{align*}
In the following section, we discuss switch scenarios between different attack actions.

\subsection{Attack patterns}\label{sec:attack}

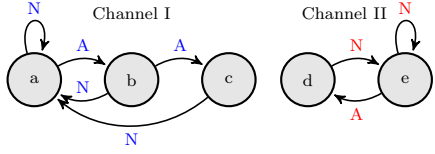
\begin{figure}[ht] 
\begin{center}
\begin{tikzpicture}[->,>=stealth',shorten >=1pt,auto,node distance=1.6cm, font=\small, semithick]
\tikzstyle{every state}=[fill=gray!20,draw=black,thick,text=black,,scale=1]
\begin{scope}[scale=0.8,transform shape]
\node[state] (A) {a};
\node (AA) [right of = A, above of = A, yshift = -.5cm] {Channel I}; 
\node[state,draw=black] (B) [right of = A] {b};
\node[state,draw=black] (C) [right of = B] {c};
\path (A) edge [bend left] node {\Blue{A}} (B);
\path (A) edge [loop above] node {\Blue{N}} (A);
\path (B) edge [bend left] node {\Blue{A}} (C);
\path (B) edge [bend left] node[above] {\Blue{N}} (A);
\path (C) edge [bend left=40] node {\Blue{N}} (A);
\begin{scope}[xshift = 4.5cm, yshift = 0cm]
\node[state] (A) {d};
\node (AA) [above of = A, xshift = .6cm, yshift = -.5cm] {Channel II};
\node[state,draw=black] (B) [right of = A] {e};
\path (A) edge [bend left] node {\Red{N}} (B);
\path (B) edge [loop above] node {\Red{N}} (B);
\path (B) edge [bend left] node {\Red{A}} (A);
\end{scope}
\end{scope}
\end{tikzpicture}
\end{center}
\caption{Two graphs representing  the attack policies carried out over channels I and II, respectively.}
\label{fig:graph_examples}
\end{figure}

We study attack policies that enable attackers to embed logic and focus on individual attack operations that are made up of two main ingredients: the targeted channel(s) and the set of logic rules (e.g., dwell-time, attack channels). We focus on logic rules that can be expressed via a regular language \cite[Chapter 2.4]{CassandrasBook2010}. The overall attack policy can thus be described by a directed labelled graph. An edge indicates a set of attack operations, each acting on a specific system signal (e.g., measurements readings, actuation) over an unsafe channel. An edge also signifies the transition of the physical process in a single time step under the set of underlying attack actions, and, thus, is associated with a specific dynamic mode (see \eqref{eq:closed_loop2}). Fig. \ref{fig:graph_examples} illustrates examples of attack policies comprising different tactics and logic rules applied to two independent channels. An edge with a label `N' denotes attack-free, nominal operation, whereas a label `A' implies FDI attack on a channel. For the example over channel I, we assume that FDI attacks cannot happen more than two consecutive time steps. Over channel II, an FDI attack has to be followed by a nominal operation for at least one time instant.
These two examples illustrate how dwell-time restrictions and admissible sequences are modeled for attack tactics.

\begin{figure}[ht] 
\begin{center}
\begin{tikzpicture}[->,>=stealth',shorten >=1pt,auto,node distance=2.5cm, font=\small, semithick]
\tikzstyle{every state}=[fill=gray!20,draw=black,thick,text=black,,scale=1]
\begin{scope}[scale=0.7,transform shape]
\node[state]                    (A)                {a};
\node[state,draw=black]         (B) [ right of =A, xshift = -.2cm, yshift = -1cm] {b};
\node[state,draw=black]         (D) [ below of =A] {d};
\node[state,draw=black]         (E) [ right of =D] {e};
\node[state,draw=black]         (C) [ right of =E, xshift = 3cm] {c};
\node[state,draw=black]         (F) [ above of = C, right of = E, yshift = .4cm] {f};
\path (A) edge [bend right = 20] node {\Blue{A}\Red{A}} (D);
\path (A) edge [loop left] node {\Blue{N}\Red{N}} (A);
\path (A) edge [bend left] node {\Blue{N}\Red{A}} (B);
\path (A) edge [out=90, in=30, out looseness=1, in looseness=1.5] node {\Blue{A}\Red{N}} (C);
\path (F) edge [bend right = 10] node {\Blue{N}\Red{N}} (A);
\path (C) edge  node {\Blue{N}\Red{A}} (B);
\path (C) edge [bend right = -10] node {\Blue{A}\Red{N}} (E);
\path (C) edge[out=70, in=30, out looseness=1, in looseness=1.5] node{\Blue{N}\Red{N}} (A);
\path (C) edge [bend right] node {\Blue{A}\Red{A}} (F);
\path (B) edge [bend left] node[above] {\Blue{N}\Red{N}} (A);
\path (B) edge [bend left] node[above,yshift = 4] {\Blue{A}\Red{N}} (C);
\path (E) edge [bend left] node {\Blue{N}\Red{A}} (B);
\path (E) edge [bend left] node {\Blue{N}\Red{N}} (A);
\path (D) edge [bend right] node {\Blue{A}\Red{N}} (E);
\end{scope}
\end{tikzpicture}
\end{center}
\caption{Graph depicting the pattern of the overall attack policy carried out over channels I and II, that combines the individual logic shown in Fig. \ref{fig:graph_examples}.} 
\label{fig:graph_product_example}
\end{figure}
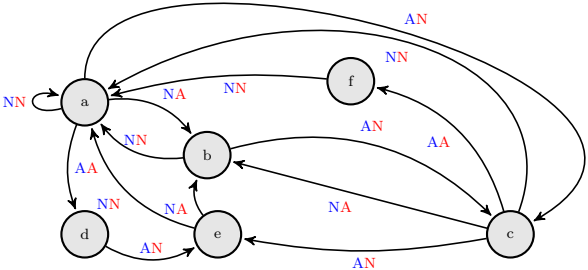

The examples in Fig. \ref{fig:graph_examples} represent constrained attack tactics carried out over a single channel. A cyber physical system, recalling Fig. \ref{fig:1}, however, can have several vulnerable points and be subject to more complex and varying attack actions. Fig. \ref{fig:graph_product_example} depicts the pattern of an overall attack policy associated with two independent attack actions taking place over two different channels lying in the same CPS. It describes all possible combinations of the allowed attacks in the two channels, and is the Kronecker product of the two individual graphs in Fig. \ref{fig:graph_examples}, defined next (see, e.g., \citep{AthanasopoulosCDC15}).
The capability to combine simple attack policies into complex ones provides a systematic approach to derive and assess a variety of attack scenarios for CPS.
\vspace{-1.35cm}

\begin{defn} Let two directed labelled graphs $\cG_1$, $\cG_2$, with sets of nodes $\cV(\cG_1) =\cV_1$, $\cV(\cG_1) =\cV_2$, respectively, and sets of edges $\cE(\cG_1) =\cE_1$, $\cE(\cG_1) =\cE_2$, respectively. Then, their Kronecker product $\cG_1 \otimes \cG_2$ has the following set of vertices and edges $\cV(\cG_1 \otimes \cG_2) = \{pq: p \in \cV_1, \; q \in \cV_2\}$ and  $\cE(\cG_1 \otimes \cG_2)  = \{ (p_1q_1,\; p_2q_2): (p_1,\;p_2) \in \cE_1,\; (q_1,\;q_2) \in \cE_2 \}$ respectively.
\end{defn}

We denote by $\mathcal{U}_c = \{i_1,\;i_2,\;\ldots,\; i_{n_u}\}$ the set of all \emph{input channels} corresponding to elements of the input vector $u(t)$, and by $\mathcal{Y}_s = \{o_1,\;o_2,\;\ldots,\; o_{n_y}\}$, the set of all \emph{output channels} corresponding to elements of the output vector $y(t)$. We denote by $\tilde{\mathcal{U}}_c \subseteq \mathcal{U}_c$ and $\tilde{\mathcal{Y}}_s \subseteq \mathcal{Y}_s$ the sets of vulnerable input and output channels, respectively. 
Let $\tilde{\cC} = \{\tilde{\mathcal{U}}_c,\;\tilde{\mathcal{Y}}_c\} = \{\tilde{c}_1,\;\ldots,\;\tilde{c}_m\}$ denote the collective set of all vulnerable input and output channels of the system, and $\tilde{\Sigma}_j$, $j = 1,\;\ldots,\;m$, be the set of attack tactics exerted on the signal(s) carried over the channel $\tilde{c}_j$. Let also $\{\cG_1(\cV_1,\;\cE_1),\;\ldots,\;\cG_m(\cV_m,\;\cE_m)\}$ be the set of graphs each representing the pattern of a distinct attack action taking place over a single channel, where $\cV_j$, and $\cE_j = \{(s,\;d,\;\sigma):\; s \in \cV_j,\; d \in \cV_j, \; \sigma \in \tilde{\Sigma}_j \}$ are the sets of nodes and labelled edges, respectively, of the graph $\cG_j$. Then, $\cG(\cV,\;\cE) = \cG_1 \otimes \cG_2 \otimes \cdots \otimes \cG_m$, represents the pattern of the overall attack policy. 

\section{Main results}\label{section:switching_sys_model}

We describe the attacked closed-loop system in the constrained switching system formalism  \citep{Athanasopoulos2018, AthanasopoulosLCSS2017}, capturing hybrid phenomena associated with attack patterns embedding logic rules. 

For example, one would like to capture the situation where an attack action is launched only for a limited period (e.g., to prevent from being identified), or switches between tactics carried out over different channels. 

\subsection{Switching-system attack modelling}

We consider a set of systems $\cP = \{P_1,\;\ldots,\;P_N\}$, where the dynamics of its mode $i$ is given in \eqref{eq:closed_loop2}. The overall attack pattern, i.e., the switch between dynamic modes is described by a directed labelled graph $\cG(\cV_{\cG},\;\cE_{\cG})$. Let the set of outgoing nodes of a node $s\in \cV_{\cG}$ be denoted by $\Out{s}{\cG}:= \{d\in \cV_{\cG}: (\exists \sigma\in \{1,\;\ldots,\;N\}: (s,\;d,\;\sigma)\in \cE_{\cG})\}$, where $\cG(\cV_{\cG},\;\cE_{\cG})$ (or simply $\cG$). We denote by $n_z$, $n_a$, and $n_h$, the augmented state, attack, and disturbance dimensions, respectively. We also consider the cartesian product of the disturbance and uncertainty sets $\cH=\cV\times \cW$. The dynamics of the overall attacked system is
\begin{align}
 z(t+1) &= Az(t) + B_{\sigma(t)}a(t) +E h(t), \label{eq:ss1}\\
\xi(t+1) & \in \Out{\xi(t)}{\cG(\cV_{\cG},\;\cE_{\cG})}, \label{eq:ss2} \\
%w(t) & \in \cW_{\sigma(t)}, \\
(z(0),\;\xi(0)) & \in \cZ \times \cV_{\cG}, \label{eq:ss3} 
\end{align}
subject to the constraints 
\begin{align}
(\xi(t),\;\xi(t+1),\;\sigma(t)) & \in \cE_{\cG}, \label{eq:con1}\\
%\sigma(t) & \in\{\sigma:(\xi(t),\;\xi(t+1),\;\sigma)\in E\}, \\
z(t) & \in \cZ, \label{eq:con2}\\
% r(t) & \in \mathcal{R}_r, \label{eq:con4} \\
h(t) & \in \mathcal{H} \label{eq:con3} \\
a(t) & \in \cA_{\sigma(t)}(z(t)), \label{eq:con4}
\end{align}
for all $t\geq 0$. We note that although the switching signal $\sigma(t)$ is a function of $z(t)$, we simply write $\sigma(t)$. System \eqref{eq:ss1}-\eqref{eq:con4} is defined in the hybrid state space\footnote{Indeed, from \eqref{eq:ss2}, \eqref{eq:ss3}, it follows that $\xi(t)\in \cV_{\cG}$, for all $t\geq 0$.} $[z^\top\; \xi]^\top\in\mathbb{R}^{n_z}\times \cV_{\cG}$. We assume that $\sigma = 1$ corresponds to the attack-free dynamics. We call \emph{nominal} the attack-free system. The stability of the autonomous system $z(t+1) = Az(t)$ is guaranteed by the stability of matrices $A_p-B_pK$, $A_p-LC_p$.
\begin{rem}
The dynamics and disturbance matrices $A$ and $E$, respectively, are identical for all modes of \eqref{eq:ss1}. Attack actions altering these matrices can also be considered in our framework.
\end{rem}
We state our assumptions next.
\begin{assum} \label{ass:Csets}
The constraint and disturbance sets $\cZ$, $\cH$, are C-sets. 
\end{assum}
\begin{assum}\label{ass:graph_nonempty}
The sets $\Out{i}{\cG(\cV_\cG,\;\cE_\cG)}$, with $i\in \cV_\cG$, are nonempty.
\end{assum}
\begin{assum}\label{ass:stealthy}
The attacker has knowledge of the system matrices $(A,\;B_i,\;C,\;E,\;D_i,\;F)$, the controller and observer gains $K$ and $L$, the state, input, and output constraint sets, $\cZ$, $\cU$, and $\cY$, and the disturbance sets $\cV$, and $\cW$. 
\end{assum}
\begin{rem}
Assumption~\ref{ass:Csets} is standard, see e.g., \citep{BlanchiniBook2015}. Note that the constraint set $\cZ$ is formed of the state, input, and output constraint sets, namely, $\cX$, $\cU$, and $\cY$. Assumption~\ref{ass:graph_nonempty} guarantees the completeness of solutions. Assumption~\ref{ass:stealthy} is standard for the construction of stealthy data poisoning attacks.
\end{rem}

The constraint \eqref{eq:con4} enforces attack stealthiness, which  is defined next. Recall $a(t) = [a_u(t)^{\top}~a_y(t)^\top]^\top$. First, we require that input poisoning attacks do not violate input constraints, i.e., $\tilde{u}(t) = u(t) + \Gamma_{\sigma(t)}^u a_u(t) \in \cU$, and let $\cU = \{u: G_u u\leq h_u\}$. We define 
\begin{align}
    \cA_{\sigma}^u(z) = \{ a_u : & (G_u)_j   \Gamma_{\sigma}^u a_u \leq (h_u)_j \nonumber \\ &+ (G_u)_j K[I_{n_x}~-I_{n_x}]z,\; j \in J_{G_u}\}.
\end{align}
We call stealthy an input attack if $a_u(t) \in \cA_{\sigma(t)}^u(z(t))$. The output poisoning attacks should respect two types of constraints, namely, the output constraints, i.e., $\tilde{y}(t) = y(t) + \Gamma_{\sigma(t)}^y a_y(t) \in \cY$, with $\cY = \{y:G_y y\leq h_y\}$, and the residual constraints, i.e., $r(t) \in \cR$, with $\cR$ as in \eqref{eq:stealthy_residual_set}. Let 
\begin{align}
    \cA_{\sigma}^y(z) = \{ a_y &: (G_y)_j\Gamma_{\sigma}^y a_y \leq (h_y)_j - \max_{w\in \cW} (G_y)_j w \nonumber \\& - (G_y)_jC_p[I_{n_x}~0_{n_x}]z,\; j \in J_{G_y}\},
\end{align}
and
\begin{align}
    \cA_{\sigma}^r(z) = \{ a_y &: (G_r)_j\Gamma_{\sigma}^y a_y \leq (h_r)_j - \max_{w\in \cW} (G_r)_j w \nonumber \\& - (G_r)_jC_{\sigma}z,\; j \in J_{G_r}\}.
\end{align}
\begin{rem}
The point-to-set maps $\cA^y_{\sigma}(z)$, $\cA^r_{\sigma}(z)$ are obtained after erosion with respect to $G_y\cW$, $G_r\cW$, respectively. Note that although the disturbance set $\cW$ is known to the attacker, the exact perturbation signal $w(t)$ is unknown. 
\end{rem}
We call stealthy an output attack if $a_y(t) \in \cA_{\sigma(t)}^y(z(t))\cap \cA_{\sigma(t)}^r(z(t))$. Then, a stealthy attack signal $a(t)$ is defined next.
\begin{defn}\label{defn:stealthiness}
The attack signal $a(t) = \begin{bmatrix}a_u(t)^\top & a_y(t)^\top \end{bmatrix}^\top$ is called stealthy if $a(t) \in \cA_{\sigma(t)}(z(t))$ where %\nikos{why not simply write the following}
\begin{align*}
\cA_{\sigma}(z)= \cA_{\sigma}^u(z)\times \left( \cA_{\sigma}^y(z) \cap \cA_{\sigma}^r(z) \right),\; \sigma=1,\;\ldots,\;N.   
\end{align*}
\end{defn}
\begin{rem}
The $H$-representation of $\cA_{\sigma}(z)$ is
\begin{align}
    \cA_{\sigma}(z) = \{ a: G_{a_{\sigma}} a \leq H_{a_{\sigma}}(z) \},
\end{align}
where $G_{a_{\sigma}}$ is a real matrix, $H_{a_{\sigma}}(z)$ is a convex piecewise affine function of $z\in \cZ$, and the inequality applies elementwise, \citep{CannonCDC15}.  
\end{rem}

Parametric convex sets are defined next.
\begin{defn}[\cite{CannonCDC15}]
Let $X\subseteq \R^n$, $Y\subseteq \R^m$, let $\cP(Y)$ denote the power set of $Y$, and $T: X \to \cP(Y)$, $X\ni s\mapsto T(s)\subset Y$ be a continuous point-to-set map. The map $T$ is called \emph{parametrically convex} if it satisfies $T(\lambda s_1 +(1-\lambda)s_2)\subseteq \lambda T(s_1) \oplus (1-\lambda) T(s_2)$ for all $s_1,\;s_2\in X$ and $0\leq \lambda \leq 1$.
\end{defn}
\begin{lem}[\citep{CannonCDC15}]
The point-to-set map $\cA_{\sigma}(z)$ is parametrically convex for all $z\in \cZ \subset \mathbb{R}^{n_z}$.
\end{lem}
\begin{lem}[\citep{CannonCDC15}] \label{lemma_compactness}
The set $\cA_{\sigma}(z)$ is pointwise compact and polytopic for all $z\in \cZ$. 
\end{lem}

\subsection{Safe set computation}
To define properly  safety for system \eqref{eq:ss1}-\eqref{eq:con4}, first, we recall the notions of multi-sets.
\begin{defn}[Multi-sets] We call multi-set a collection of sets $\{\cS^i\}_{i\in \cV_{\cG}}$, with $\cS^i \subset \R^{n_z}$, $i \in \cV_{\cG}$.
\end{defn}
\begin{defn}[Invariance] The multi-set $\{\cS^i\}_{i\in \cV_{\cG}}$ is an invariant multi-set with respect to \eqref{eq:ss1}-\eqref{eq:con4} if $z(0) \in \cS^{\xi(0)}$ implies $z(t) \in \cS^{\xi(t)}$ for all $t\geq 0$, $\xi(0) \in \cV_{\cG}$, and $\sigma(t)$ satisfying \eqref{eq:con1}. If, additionally, $\cS^i \subset \cZ$, $i \in \cV_{\cG}$, then, $\{\cS^i\}_{i\in \cV_{\cG}}$ is called an \emph{admissible invariant multi-set} with respect to \eqref{eq:ss1}-\eqref{eq:con4}. The multi-set $\{\cS_M^i\}_{i \in \cV_{\cG}}$ is the \emph{maximal admissible invariant multi-set} if for any admissible invariant multi-set $\{\cS^i\}_{i \in \cV_{\cG}}$, it holds that $\cS^i \subseteq \cS_M^i$, $i \in \cV_{\cG}$. The invariant multi-set $\{\cS_m^i\}_{i\in \cV_{\cG}}$ is the \emph{minimal invariant multi-set} if for any invariant multi-set $\{\cS^i\}_{i\in \cV_{\cG}}$ it holds $\cS_m^i\subseteq \cS^i$ , $ i \in \cV_{\cG}$. 
\end{defn}

\begin{defn}[Safety]\label{def:safety}
A set $\cS_{\cV_{\cG}}\subset \mathbb{R}^{n_z}$ is safe with respect to system \eqref{eq:ss1}-\eqref{eq:con4} and the set of nodes $\cV_{\cG}$ if $(z(0),\;\xi(0)) \in \cS_{\cV_{\cG}} \times \cV_{\cG}$, implies $z(t) \in \cZ$, $t\geq 0$. 
\end{defn}

Consider the system \eqref{eq:ss1}-\eqref{eq:ss3} and a given switching signal $\sigma\in \{1,\;\ldots,\;N\}$. We define the \emph{one-step forward reachability map} 
\begin{align}
\Phi(\sigma,\cS)= \big\{ y: \big( \exists (z,a,h)&\in\cS\times \cA_\sigma(z)\times \cH: \nonumber\\& 
y=Az+B_\sigma a+E h \big) \big\}
\end{align}
and the \emph{one-step backward reachability map} 
\begin{align}
    \Psi(\sigma,\; \cS) = \left\{z: \left(A_{\sigma}z  \oplus B_{\sigma} \cA_{\sigma}(z)  \oplus E_{\sigma} \cH\right)  \in \cS  \right\}.\label{eq:back_reach_map}
\end{align}
The minimal-invariant multi-set is characterised next. 
\begin{prop}
Consider the forward reachability multi-set sequence $\{\cF_l^i\}_{i\in \cV_{\cG}}$, $l\geq 0$, with
\begin{align}
    \cF_0^i & = \{0\}, \; i \in \cV_{\cG}, \\
    \cF_{l+1}^i & = \cup_{(s,\;i,\;\sigma)\in \cE_{\cG}}\Phi(\sigma,\; \cF_l^s), \; i \in \cV_{\cG}. \label{eq_forward} 
\end{align}
The minimal invariant multi-set $\{\cS_m^i\}_{i\in \cV_{\cG}}$ with respect to \eqref{eq:ss1}-\eqref{eq:con4} is unique and equal to $\cS_m^i = \lim_{j \to \infty}\cF_j^i$, $i \in \cV_{\cG}$. 
\end{prop}
\begin{pf}
The proof follows the same steps as in \cite[Theorem 1]{AthanasopoulosLCSS2017}. The only difference here concerns the involvement of the state-dependent set $\cA_\sigma(z)$ in the multi-set sequence update \eqref{eq_forward}, which however has no effect in the steps of the proof since the sets $\cA_\sigma(\cZ)=\cup_{z\in\cZ}\cA_\sigma(z)$, $\sigma=1,\;\ldots,\;N$ are compact by the parametric convexity and compactness  of the set $\cA_\sigma(z)$ (Lemma~\ref{lemma_compactness}), and compactness assumption of $\cZ$, in a similar way as outlined also in the proof of \cite[Lemma 3.2]{CannonCDC15}. \qed
\end{pf}

Without loss of generality, we assume the following. 
\begin{assum}\label{ass:non_empty_safe_set}
Let $\{\cS_m^i\}_{i\in \cV_{\cG}}$ be the minimal-invariant multi-set with respect to \eqref{eq:ss1}-\eqref{eq:con4}. Then, $\cS_m^i \subset \cZ$, $i \in \cV_{\cG}$. 
\end{assum}

We consider the backward reachability multi-set sequence $\{\cB_l^i\}_{i \in \cV_{\cG}}$, where
\begin{align}
    \cB_0^i &= \cZ, \ \ i\in\cV_\cG, \label{eq:back1} \\
    \cB_{l+1}^i & = (\cB_0^i\cap_{(i,\;d,\;\sigma)\in \cE_{\cG}}\Psi(\sigma,\;\cB_l^d)),\; i \in \cV_{\cG}. \label{eq:back2}
\end{align}
The $l$th term of the multi-set sequence \eqref{eq:back1}-\eqref{eq:back2} contains the initial conditions $(z(0),\;\xi(0))$ which satisfy the state constraints for at least the first $l$ instants. Intuitively, each set $\cB_{l+1}^i$, $l \geq 0$, $i \in \cV_{\cG}$, contains the set of states in the state constraint set $\cZ$ that can be stirred to $\cB_l^d$ via the dynamics $\sigma$, where $d$ is any outgoing node of $i$, \citep{Athanasopoulos2018}. 

\begin{rem}
Let $\cB_l^d = \{z: (G_l^d)_j z \leq (g_l^d)_j,\; j \in J_{G_l^d}\}$. Then, the backward reachability map $\Psi(\sigma,\;\cB_l^d))$ is computed by enforcing the constraint 
\begin{align}
    (G_l^d)_j \left( A z + B_{\sigma} a+ E h \ \right) \leq (g_l^d)_j, \; \forall a \in \cA_{\sigma}(z), \; \forall h \in \cH,    
\end{align}
for all $j \in I_{G_l^d}$, or 
\begin{align}
    (G_l^d)_j  A z \leq   (g_l^d)_j - \max_{a\in \cA_{\sigma}(z)} (G_l^d)_j B_{\sigma} a -(G_l^d)_j E h_j^\ast,    \label{eq:MPLP}
\end{align}
for all $j \in  I_{G_l^d}$, where $h_j^\ast = \text{argmax}_{h\in \cH}(G_l^d)_j E h$. To compute the set induced by \eqref{eq:MPLP}, we need to solve $\max_{a\in \cA_{\sigma}(z)} (G_l^d)_j B_{\sigma} a$ which is a multi-parametric linear program (mpLP) with optimisers being affine functions of $z$. Solutions can be obtained, e.g., using off-the-shelf  multi-parametric programming software. Typically, the set of \emph{parameters} (here, the constraint set $\cZ$) is divided into \emph{critical regions}. Throughout a critical region, the optimality conditions derived from the KKT conditions are invariant, \citep{Borrelli2003}. In each critical region, the solution is expressed as an affine function of $z$ resulting in a new inequality describing the set $\cB_{l+1}^i$ with $(i,\;d,\;\sigma)\in\cE_{\cG}$. The multi-parametric solution is consistent with the approach in \cite{CannonCDC15}, where a similar manipulation is carried out, however, with the assumption therein that state-dependent sets have an explicit vertex representation.
\end{rem}

\begin{thm}\label{thm:max_multi}
Consider the system \eqref{eq:ss1}-\eqref{eq:con4} and the sequence \eqref{eq:back1}-\eqref{eq:back2}. Let $\{\cS_m^i\}_{i \in \cV_{\cG}}$ be the minimal invariant multi-set. Then, there is a finite $\bar{k}\geq 0$ such that $\cB_{\bar{k}+1}^i = \cB_{\bar{k}}^i$, $i \in \cV_{\cG}$. Moreover, $\{\cB_{\bar{k}}^i\}_{i \in \cV_{\cG}}$ is the maximal admissible invariant multi-set.
\end{thm}
\begin{pf}
The proof follows similar steps as in \cite[Theorem 3]{AthanasopoulosLCSS2017}. The difference in this paper is the involvement of the state-dependent set $\cA_\sigma(z)$ in the backward reachability map \eqref{eq:back_reach_map} and, consequently, in the multi-set sequence \eqref{eq:back1}-\eqref{eq:back2}. Nevertheless, this does not affect the proof development as the key requirement, namely that the set $\cA_\sigma(\cZ)$, $\sigma=1,\;\ldots,\;N$ are bounded, holds. \qed
\end{pf}

From Definition \ref{def:safety} and Theorem \ref{thm:max_multi}, the maximal safe set of \eqref{eq:ss1}-\eqref{eq:con4} is derived in the following corollary. 
\begin{cor}[\citep{AthanasopoulosLCSS2017}]\label{cor:max_safe_set} Let the maximal invariant multi-set with respect to \eqref{eq:ss1}-\eqref{eq:con4} be $\cB_M^i$, $i\in \cV_{\cG}$. The maximal safe set $\cS_{\cV_{\cG}}$ of \eqref{eq:ss1}-\eqref{eq:con4} with node set $\cV_{\cG}$ is $\cS_{\cV_{\cG}} = \cap_{i\in \cV_{\cG}}\cB_M^i$. 
\end{cor}

\begin{rem}
Assumption \ref{ass:non_empty_safe_set} can be lifted. In fact, if the inclusion $\cS_m^i\subset\cZ$, $i\in \cV_{\cG}$ does not hold, the multi-set sequence \eqref{eq:back1}-\eqref{eq:back2} converges to the empty set and, thus, the maximal safe set is empty indicating an attack with maximum impact. Convergence to empty set can be identified in finite time by checking $\cB_l^i\subset \interior(\cZ)$, $i\in \cV_{\cG}$, $l\geq 0$, \cite{BlanchiniBook2015}. 
\end{rem}

\subsection{Impact metrics}\label{section:metrics}

The maximal safe set $\cS_{\cV_{\cG}}$ in Corollary \ref{cor:max_safe_set} provides a security measure of the system under attack. To construct scalar security indices, we propose two complementary impact metrics related to the Lebesgue measure, and the Minkowski distance between sets. These are defined next.

\begin{defn}
The outer Lebesgue measure of the set $\cS\subset \R^n$ is
\begin{equation}
    \textnormal{vol}(\cS)=\inf\left\{ \sum_{j=1}^\infty\vol(\cR_j): \cS \subset \cup_{j=1}^\infty \cR_j \right\},
\end{equation}
where the infimum is taken over all countable collections of rectangles $\cR_j = [a_1^j,\;b_1^j]\times [a_2^j,\;b_2^j]\times \ldots\times [a_n^j,\;b_n^j]\in\R^n$, with $ a_l^j\leq b_l^j\in \R$, whose union contains $\cS$.
\end{defn}
\begin{defn}
Let $\cS_1\subset \R^n$, $\cS_2\subset \R^n$ be two C-sets. The Minkowski distance between $\cS_1$ and $\cS_2$ is defined as 
\begin{equation}
    \mu(\cS_1,\;\cS_2) = \max\{\lambda: \lambda \cS_1 \subseteq \cS_2\}.
\end{equation}
\end{defn}

Denote the constrained switching system \eqref{eq:ss1}-\eqref{eq:ss3} by $\cP$, and let $\cZ$, $\cH$, and $\cA$ be the constraint, disturbance, and attack sets, respectively. Let $\cS_{\cV_{\cG}}$ be the maximal safe set of \eqref{eq:ss1}-\eqref{eq:con4} and $\cS^0$ be the maximal safe set of the attack-free system. Then, we can define
\begin{equation}
    \mathcal{I}_1(\cP,\;\cZ,\;\cH,\;\cA) = \frac{\vol(\cS^0) - \vol(\cS_{\cV_{\cG}})}{\vol(\cS^0)},
\end{equation}
and
\begin{equation}
    \mathcal{I}_2(\cP,\;\cZ,\;\cH,\;\cA)= 1 - \mu(\cS^0,\;\cS_{\cV_{\cG}}),
\end{equation}
as two safety metrics of \eqref{eq:ss1}-\eqref{eq:con4}.

Clearly, $\cS_{\cV_{\cG}} \subseteq \cS^0$, and consequently $\vol(\cS_{\cV_{\cG}}) \leq \vol(\cS^0)$ and $\mu(\cS^0,\;\cS_{\cV_{\cG}})\in[0,1]$, thus, $0\leq \mathcal{I}_i \leq 1$, $i=1,2$. A metric near zero indicates an attack with little impact whereas a metric almost equal to one translate an impactful attack inducing a small safe set.  By computing $\mathcal{I}_1$, $\mathcal{I}_2$, for different attack scenarios and patterns, the safety of a CPS is evaluated, and its critical components are assessed in terms of their vulnerability to malicious exogenous inputs. Metric $\mathcal{I}_1$ provides an index of the size of a safe set the shape of which is not critical to the metric calculation. Metric $\mathcal{I}_2$, however, is sensitive to the shape of the safe set (e.g., its skewness). These are exemplified in the following section.

\begin{rem}
As the multi-set sequence \eqref{eq:back1}-\eqref{eq:back2} is monotonically nonincreasing (nested), any intersection $\cap_{i\in \cV_{\cG}}\cB_l^i$, with $l\geq 0$, will provide an underapproximation of the associated metric.
\end{rem}

\section{Numerical example}\label{sec:example}

\begin{figure}[ht]
    \centering
    \includegraphics[width =.95\linewidth]{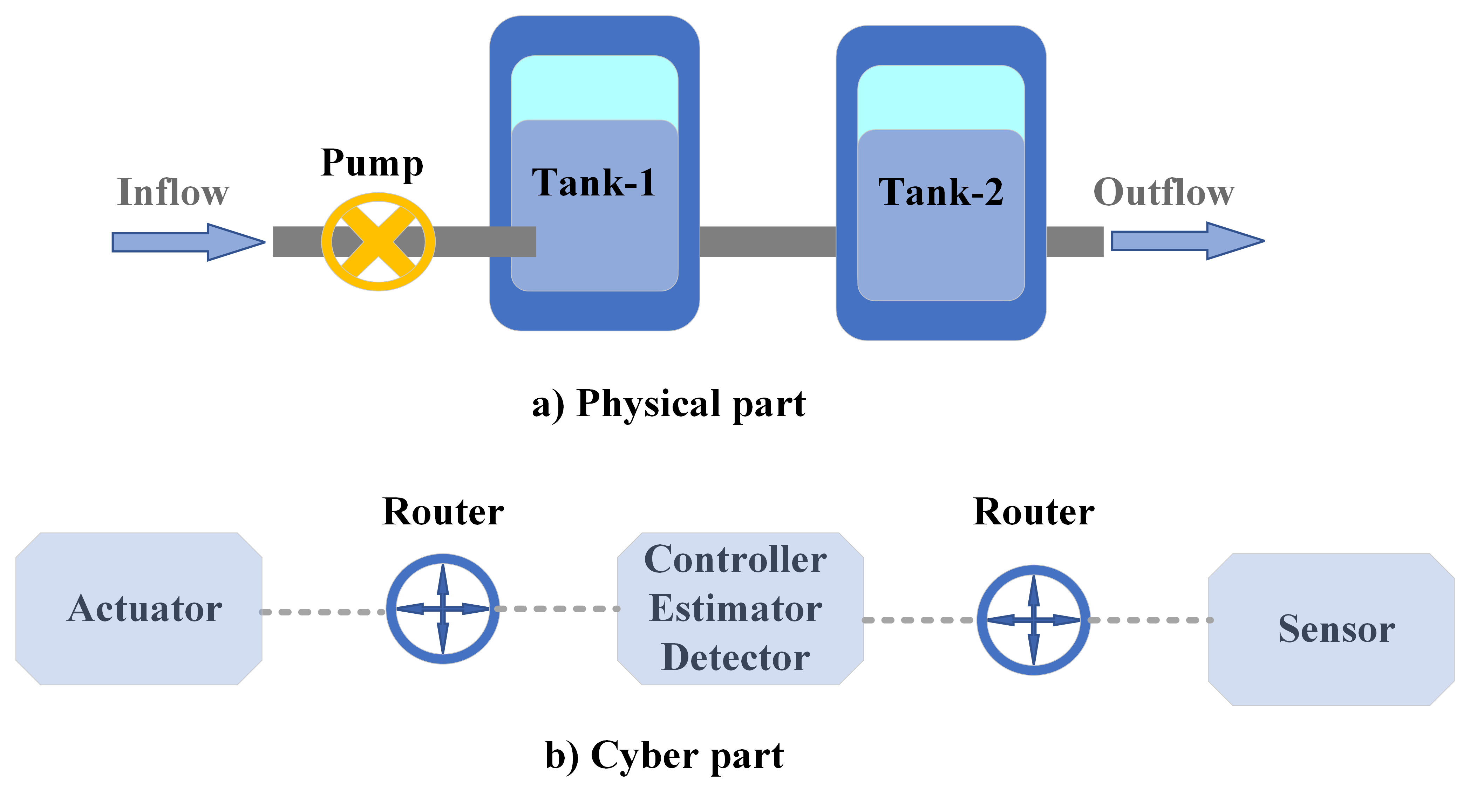}
    \caption{Physical and cyber parts of a two-tank system.}
    \label{fig:ex1}
\end{figure}

We consider a two-tank system with tanks connected as shown in Fig. \ref{fig:ex1}. The states are the liquid levels in the two tanks, denoted by $x = [x_1\;x_2]^\top$, and the input, denoted by $u$, is the flow rate of the pump. The control objective is to maintain the liquid levels at an operating point by regulating the flow rate $u$. The dynamics of the plant is $x(t+1) = Ax(t) + Bu(t) + v(t)$, where $A = \begin{bmatrix}
0.9 & 0.1\\ 0.1 & 0.8
\end{bmatrix}$, $B = \begin{bmatrix}
0.1\\0
\end{bmatrix}$, and $\|v(t)\|_{\infty} \leq 0.01$. 
The plant is equipped with a sensor measuring the liquid level of a tank, an observer estimating the system state, and a detector monitoring adversarial presence. The detector raises an alarm unless the residual $|r(t)|\leq 0.01$. The output of the system $y(t) = Cx(t) + w(t)$ is the sensor's readings, with $C = \begin{bmatrix}
1 & 0
\end{bmatrix}$ when the sensor is placed in Tank-1, and $C = \begin{bmatrix}
0 & 1
\end{bmatrix}$ when the sensor is placed in Tank-2, and $\|w(t)\|_{\infty}\leq 0.01$ accounting for noise. We show that attacks poisoning readings of the sensor placed in Tank-2 have less impact on the system safety. We consider the following attack scenarios; a) poisoning attacks on the sensor's readings, poisoning attacks on the actuation signal $u(t)$, and poisoning attacks both on sensor's readings and the actuation signal $u(t)$. The controller and observer gains $K$, $L$, are designed such that the eigenvalues of $A-BK$ and $A-LC$ are $(0.7,\;0.8)$ and $(0.86,\;0.001)$, respectively. The operating point is $x^* = [2\;1]^\top$ with $u^* = 1$. The state constraints are $1\leq x_1(t) \leq 3$, $0 \leq x_2(t) \leq 2$, and the input constraint is $0\leq u(t) \leq 2$. The attacked output signal is $\tilde{y}(t) = y(t) + a_y(t)$, and the attacked input signal is $\tilde{u}(t) = u(t) + a_u(t)$. The attack signals $a_y(t)$, $a_u(t)$, are consistent with the stealthiness Definition \ref{defn:stealthiness}. We additionally consider that the attack signals are bounded with lower and upper limits shown in Table \ref{table:attack_actions}. We associate with an individual attack action two parameters, namely, the maximum and minimum dwell times, denoted by $N_{\text{max}}$, $N_{\text{min}}$, respectively. The former is an upper bound on the time over which an attack is carried out after an attack-free operation, whereas the latter indicates a lower bound on the time of an attack-free operation after the course of a poisoning attack. The patterns of individual attack actions in terms of $N_{\text{max}}$ and $N_{\text{min}}$ are listed in Table \ref{table:attack_actions}. 
\begin{table}[ht]
\begin{center}
\caption{}\label{table:attack_actions}
\begin{tabular}{c|c|c}
    Vulnerable point & Attack bounds & Pattern \\\hline
    Sensor & $-0.05\leq a_y \leq 0.05$ & $N_{\text{max}} = N_{\text{min}} + 1$ \\\hline
    Actuator & $-0.01\leq a_y \leq 0.01$ & $N_{\text{max}} = N_{\text{min}} - 1$ \\\hline
\end{tabular}
\end{center}
\end{table}

\begin{figure}[ht]
    \centering
    \includegraphics[width=9cm, height = 6.cm]{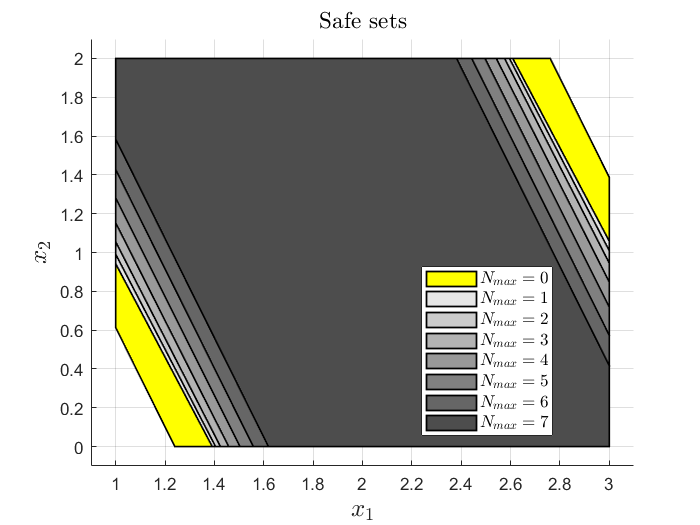}
    \caption{Safe sets of the system under attack on sensor readings associated with the liquid level of Tank-2 for various dwell time options. The plots are projections of safe sets onto $\R^2$ for $e = x - \hat{x} =0$.}
    \label{fig:safe_sets}
\end{figure}

\begin{figure}[ht]
    \centering
    \includegraphics[width=9cm, height = 6.cm]{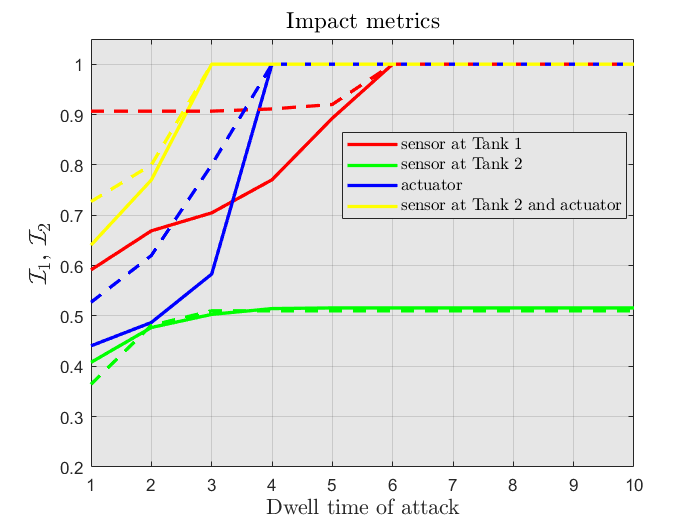}
    \caption{Impact metrics with respect to maximum dwell time. Solid lines correspond to $\mathcal{I}_1$ and dashed lines to $\mathcal{I}_2$.}
    \label{fig:metrics}
\end{figure}

We wish to quantify the safety of the plant under the attack scenarios considered. In Fig. \ref{fig:safe_sets}, we compute the safe set of the system when the reading of the sensor placed in Tank-2 is under a poisoning attack. The safe set of the attack-free dynamics is illustrated in yellow. The sets in gray are safe state regions of the system under attack for the associated dwell-time specifications. Clearly, the safe region shrinks as $N_{\text{max}}$ grows indicating safety degradation. In Fig. \ref{fig:metrics}, we compute the safety metrics $\mathcal{I}_1$, $\mathcal{I}_2$, introduced in Section \ref{section:metrics}, for all attack scenarios considered. We show that attacks poisoning the actuation signal have a major effect on the system safety, highlighting the consistency of our approach with established results. From Fig. \ref{fig:metrics}, we also conclude that a sensor placed at Tank-2 results in a less vulnerable plant preventing a safe set from collapsing to the empty set as $N_{\text{max}}$ grows. Metrics equal to one indicate an empty safe set, i.e., there is no safe initial condition close to the equilibrium point under the attack pattern considered. 


\balance

\section{Conclusion}\label{sec:conclusion}

We proposed a new approach to modelling attack scenarios in cyber-physical systems. We define a cyber-physical system under attack as a constrained switching system embedding the dynamics of the plant, the attack patterns, and the attack detection scheme. We show that our method is compatible with established results in the analysis of constrained switching systems, if additionally state dependent exogenous signals are considered, allowing us to quantify the impact of cyber attacks on the safety properties of the system. By calculating the maximal safe set of the underlying constrained switching system, we characterise system safety as an asymptotic property. Two complementary scalar metrics for security assessment are also introduced. Our switching-system approach to attack modelling is consistent with various additional attack types standard in the context of CPSs. This will be presented in a future work.

\bibliography{IFAC2023}

\end{document}